\documentclass[12pt,a4paper]{article}
\usepackage{amssymb,amsmath,amsthm}
\usepackage{natbib}
\theoremstyle{plain}

\theoremstyle{definition}

\numberwithin{equation}{section}
\usepackage{fullpage}

\usepackage{cite}

\usepackage{graphicx}

\begin{document}

\title{A note on heterogeneous beliefs with CRRA utilities}

\author{ A.A. Brown\thanks{Wilberforce Road, Cambridge CB3 0WB, UK (phone = +44 1223 337969 , email = A.A.Brown@statslab.cam.ac.uk)}  \\ \small Statistical Laboratory, \\  \small University of Cambridge  }

\date{July 2009}

\maketitle

\abstract{This note will extend the research presented in \citet{BrownRogersDiverseBeliefs} to the case of CRRA agents.  We consider the model outlined in that paper in which agents had diverse beliefs about the dividends produced by a risky asset.  We now assume that the agents all have CRRA utility, with some integer coefficient of relative risk aversion.  This is a generalisation of \citet{BrownRogersDiverseBeliefs} which considered logarithmic agents.  We derive expressions for the state price density, riskless rate, stock price and wealths of the agents.  This sheds light on the effects of risk aversion in an equilibrium with diverse beliefs.  }

\section{Introduction}

This paper will explore heterogeneous beliefs in the case in which agents have CRRA utility.  \citet{BrownRogersDiverseBeliefs} explained the general theory of how to incorporate diverse beliefs into a multi-agent equilibrium model.  That paper then specialised to the explore the example in which agents had logarithmic utility.  It was shown that calculations were very tractable and expressions for the stock price, riskless rates and wealths were all derived.   That paper also discussed the large literature on heterogeneous beliefs; readers interested in other literature in the area should refer to that paper. 

The purpose of this note is then to illustrate that the case of CRRA agents with integer coefficient of relative risk aversion is no harder to deal with than the logarithmic case.  Although the expressions we obtain are more complicated, the means by which we get these expressions are the same as in the logarithmic case.   The expressions that we derive will allow us to explore the effects of risk aversion on quantities such as stock price, riskless rates and wealths. \citet{BrownRogersDiverseBeliefs} were unable to do this, since their example used logarithmic agents.  This note therefore gives an important generalisation to previous work.

The structure of this note is as follows.  We first give a brief overview of the setup.  We then use this to obtain the state price density.  This can then be used to work out the wealth of the agents and the stock price.  We then explore the riskless rate and the stock price dynamics.

\section{CRRA Agents}

\subsection{The Setup}

The setup is exactly the same as in \citet{BrownRogersDiverseBeliefs} and readers are assumed to be familiar with this paper.  The only difference is that agents now have CRRA utility, with some integer coefficient of relative risk aversion, assumed to be greater than 1.   

As before there is a single risky asset producing dividends continuously in time.  There is also a riskless asset in zero net supply.  Agents all optimise their expected utility of consumption, but they each work under their own measure. Agent $j$ is assumed to work under measure $\mathbb{P}^{j}$.  There is also a reference probability measure, which is denoted by $\mathbb{P}^{0}$.  

We also assume that the agents have CRRA utility, with integer coefficient of relative risk aversion.  Thus, we take:
\begin{align*}
U_{j} (t, x) = e^{-\rho_{j} t} \dfrac{x^{1-R}}{1-R}
\end{align*}
for some integer $R \in \{2,3,...\}$.  The case in which agents have a constant of relative risk aversion equal to one is the case of logarithmic agents and was considered in \citet{BrownRogersDiverseBeliefs}.

\subsection{The State Price Density}

The paper of \citet{BrownRogersDiverseBeliefs} tells us that the state price density satisfies:
\begin{align*}
\zeta_{t} \nu_{j} = U_{j}' (t, c_{t}^{j}) \Lambda_{t}^{j}
\end{align*}
where $\zeta_{t}$ is the state price density, $\nu_{j}$ is a constant particular to agent $j$ and $\Lambda_{t}^{j}$ is the change of measure martingale for agent $j$  \footnote{See \citet{BrownRogersDiverseBeliefs} for further explanation and derivation of this expression.}.  Substituting for our expression for $U_{j}$ and rearranging gives: 
\begin{align*}
c_{t}^{j} = \zeta_{t}^{-1/R} \left(\dfrac{e^{-\rho_{j} t} \Lambda_{t}^{j}}{\nu_{j}}  \right)^{1/R}
\end{align*}
Since the total consumption must equal the total output of the economy, we obtain:
\begin{align}
\zeta_{t} = \delta_{t}^{-R}  \left( \sum_{i} \left(\dfrac{e^{-\rho_{i} t} \Lambda_{t}^{i}}{\nu_{j}}  \right)^{1/R} \right)^{R}  \label{CRRASPD}
\end{align}
where $\delta_{t}$ is the dividend process produced by the risky asset.  Using this state price density we can then compute the wealth of agents, the price of the risky asset and the riskless rate.

\subsection{The Wealth of Agents}

The wealth of agent $j$ is given by:
\begin{align}
w_{t}^{j} & = \mathbb{E}_{t}^{0} \left[\int_{t}^{\infty} \dfrac{ c_{u}^{j} \zeta_{u}}{\zeta_{t}} du  \right] \notag \\
& = \zeta_{t}^{-1} \mathbb{E}_{t}^{0} \left[\int_{t}^{\infty} \left(\Lambda_{u}^{j} e^{-\rho_{j} u}/\nu_{j} \right)^{1/R} \delta_{u}^{1-R} \left(\sum_{i} \left(e^{-\rho_{i} u} \Lambda_{u}^{i} / \nu_{i}  \right)^{1/R}  \right)^{R-1}  du \right]  \label{wealthexpv1}
\end{align}
Let $\gamma_{i} = \log \nu_{i}$.  We will assume that
\begin{align}
d\Lambda_{t}^{j} = \Lambda_{t}^{j} \alpha_{j} dX_{t}
\end{align}
where $(X_{t})_{t \ge 0}$ is a standard Brownian motion under the reference measure $\mathbb{P}^{0}$.   Then we have that
\begin{align*}
 \left(e^{-\rho_{i} u} \Lambda_{u}^{i} / \nu_{i}  \right)^{1/R} = \exp \{ -\dfrac{(\rho_{i} u + \gamma_{i})}{R} + \dfrac{\alpha_{i}}{R} X_{u} - \dfrac{1}{2} \dfrac{\alpha_{i}^{2}}{R} u   \}
\end{align*}
Since $R$ is an integer\footnote{Note that we have not needed the assumption that $R$ is an integer until this point}, we may perform a multinomial expansion on the final term in expression (\ref{wealthexpv1}) to give:  
\begin{multline*}
\left(\sum_{i} \left(e^{-\rho_{i} u} \Lambda_{u}^{i} / \nu_{i}  \right)^{1/R}  \right)^{R-1} = \sum_{|\beta| = R-1} \binom{R-1}{\beta} \exp\{ -\dfrac{(\rho \cdot \beta u + \gamma \cdot \beta )}{R} \\ 
+ \dfrac{\alpha \cdot \beta}{R} X_{u} - \dfrac{1}{2} \dfrac{\alpha^{2} \cdot \beta}{R} u \}
\end{multline*} 
Here,  $\alpha$ and $\rho$ denote the vectors of all the different $\alpha_{j}'s$ and $\rho_{j}$'s respectively. The $\binom{R-1}{\beta}$ term denotes the multinomial coefficient $\binom{R-1}{\beta_{1}, \beta_{2},..., \beta_{J}}$.  The summation is over all vectors $\beta$ of length $J$, with each entry taking values in $\{0,1,...,R-1 \}$ and such that $\sum_{i} \beta_{i} = R-1$.   

As in \citet{BrownRogersDiverseBeliefs}, we assume that the dividend at time $u$ is given by:
\begin{align*}
\delta_{u}=\delta_{0} \exp \{\sigma X_{u} + \left( \alpha^{*} \sigma - \dfrac{\sigma^{2}}{2}\right) u  \}
\end{align*}
Putting this together yields:
\begin{multline*}
w_{t}^{j} = \delta_{0}^{1-R} \zeta_{t}^{-1} \int_{t}^{\infty} \sum_{|\beta| = R-1} \binom{R-1}{\beta} \mathbb{E}_{t}^{0} \big[\exp \{- \dfrac{ \rho_{j} u + \gamma_{j}}{R} + \dfrac{\alpha_{j}}{R} X_{u} - \dfrac{1}{2} \dfrac{\alpha_{j}^{2}}{R} u   \\
+ (1-R) \sigma X_{u} + (\alpha^{*} \sigma - \dfrac{\sigma^{2}}{2} ) (1-R) u - \dfrac{ \rho \cdot \beta u + \gamma \cdot \beta }{R} + \dfrac{\alpha \cdot \beta}{R} X_{u} - \dfrac{1}{2} \dfrac{\alpha^{2} \cdot \beta}{R} u \} \big] du
\end{multline*}
It is then straightforward to compute the conditional expectation.  We obtain:
\begin{multline*}
w_{t}^{j} = \delta_{0}^{1-R} \zeta_{t}^{-1} \int_{t}^{\infty} \sum_{|\beta| = R-1} \binom{R-1}{\beta} \exp\{ \left(\dfrac{\alpha_{j} + \alpha \cdot \beta}{R} + (1-R) \sigma \right) X_{t} \\
+ \dfrac{1}{2} \left(\dfrac{\alpha_{j} + \alpha \cdot \beta}{R} + (1-R) \sigma \right)^{2} (u-t) \}  \\
\exp\{ - \dfrac{\gamma_{j} + \gamma \cdot \beta}{R} - \left[\dfrac{\rho_{j} + \rho \cdot \beta}{R} + \dfrac{\alpha_{j}^{2} + \alpha^{2} \cdot \beta}{2 R} + \left(\dfrac{\sigma^{2}}{2} - \alpha^{*} \sigma \right) (1-R)   \right] u \}  du
\end{multline*}
Finally, we may compute the integral to obtain:
\begin{multline*}
w_{t}^{j} = \delta_{0}^{1-R} \zeta_{t}^{-1} \sum_{|\beta| = R-1} \binom{R-1}{\beta} \exp\{ \left(\dfrac{\alpha_{j} + \alpha \cdot \beta}{R} + (1-R) \sigma \right) X_{t}  - \dfrac{\gamma_{j} + \gamma \cdot \beta}{R}  \}  \\
\exp\{  - \left[\dfrac{\rho_{j} + \rho \cdot \beta}{R} + \dfrac{\alpha_{j}^{2} + \alpha^{2} \cdot \beta}{2 R} + \left(\dfrac{\sigma^{2}}{2} - \alpha^{*}\sigma  \right) (1-R)   \right] t \} \\
  \left[\dfrac{\rho_{j} + \rho \cdot \beta}{R} + \dfrac{\alpha_{j}^{2} + \alpha^{2} \cdot \beta}{2 R} + \left(\dfrac{\sigma^{2}}{2} - \alpha^{*}\sigma \right) (1-R) - \dfrac{1}{2} \left(\dfrac{\alpha_{j} + \alpha \cdot \beta}{R} + (1-R) \sigma \right)^{2}  \right]^{-1}
\end{multline*}
where we have assumed that the final line of this expression is positive for all $\beta$, in order that the integral is finite\footnote{A sufficient condition for the integral to be finite is that 
\begin{align*}
\min_{i} (\rho_{i} + \dfrac{\alpha_{i}^{2}}{2}) + \left(\dfrac{\sigma^{2}}{2} - \alpha^{*}\sigma \right) (1-R) - \dfrac{1}{2} \max_{i} \{ ((R-1) \sigma -  \alpha_{i})^{2} \} \ge 0  
\end{align*}}. We may also use our expression for $\delta_{t}$ to obtain:
\begin{multline}
w_{t}^{j} = \delta_{t}^{1-R} \zeta_{t}^{-1} \sum_{|\beta| = R-1} \binom{R-1}{\beta} \exp\{ \dfrac{\alpha_{j} + \alpha \cdot \beta}{R}  X_{t}  - \dfrac{\gamma_{j} + \gamma \cdot \beta}{R}  - \left[\dfrac{\rho_{j} + \rho \cdot \beta}{R} + \dfrac{\alpha_{j}^{2} + \alpha^{2} \cdot \beta}{2 R}  \right] t \} \\
  \left[\dfrac{\rho_{j} + \rho \cdot \beta}{R} + \dfrac{\alpha_{j}^{2} + \alpha^{2} \cdot \beta}{2 R} + \left(\dfrac{\sigma^{2}}{2} - \alpha^{*} \sigma\right) (1-R) - \dfrac{1}{2} \left(\dfrac{\alpha_{j} + \alpha \cdot \beta}{R} + (1-R) \sigma \right)^{2}  \right]^{-1}  \label{CRRAWealth}
\end{multline}

\subsection{The Stock Price}

To compute the stock price, we may simply sum the wealth (or compute the net present value of future dividends directly) to obtain:
\begin{multline}
S_{t} = \delta_{t}^{1-R} \zeta_{t}^{-1} \sum_{|\beta| = R} \binom{R}{\beta} \exp\{ \dfrac{\alpha \cdot \beta}{R}  X_{t}  - \dfrac{\gamma \cdot \beta}{R}  - \left[\dfrac{\rho \cdot \beta}{R} + \dfrac{\alpha^{2} \cdot \beta}{2 R}  \right] t \} \\
  \left[\dfrac{ \rho \cdot \beta}{R} + \dfrac{ \alpha^{2} \cdot \beta}{2 R} + \left(\dfrac{\sigma^{2}}{2} - \alpha^{*} \sigma\right) (1-R) - \dfrac{1}{2} \left(\dfrac{ \alpha \cdot \beta}{R} + (1-R) \sigma \right)^{2}  \right]^{-1} \label{CRRASP1}
\end{multline}
Using the expression for the state price density given in (\ref{CRRASPD}), we obtain the following expression for the price-dividend ratio:
\begin{multline*}
S_{t}/\delta_{t} = \left( \sum_{i} \exp\{-\dfrac{\rho_{i} }{R}t - \dfrac{\gamma_{i}}{R} + \dfrac{\alpha_{i} }{R} X_{t} - \dfrac{\alpha_{i}^{2} }{2 R}t \}  \right)^{-R} \\
\sum_{|\beta| = R} \binom{R}{\beta} \frac{\exp\{ \dfrac{\alpha \cdot \beta}{R}  X_{t}  - \dfrac{\gamma \cdot \beta}{R}  - \left[\dfrac{\rho \cdot \beta}{R} + \dfrac{\alpha^{2} \cdot \beta}{2 R}  \right] t \}}{  \left[\dfrac{ \rho \cdot \beta}{R} + \dfrac{ \alpha^{2} \cdot \beta}{2 R} + \left(\dfrac{\sigma^{2}}{2} - \alpha^{*} \sigma\right) (1-R) - \dfrac{1}{2} \left(\dfrac{ \alpha \cdot \beta}{R} + (1-R) \sigma \right)^{2}  \right] }
\end{multline*}

\paragraph{Remarks}
If all the agents have the same discount factor and the agents have logarithmic utility, then \citet{BrownRogersDiverseBeliefs} showed that the price-dividend ratio was simply  $\rho_{1}^{-1}$.  In particular, this implied that the volatility of the stock was the same as the volatility of the dividend process.  This was quite unsatisfactory, since it implied that the diverse beliefs only had an impact on the stock volatility if the agents had different discount factors.

The case of CRRA agents with integer coefficient of risk aversion ($R \ge 2$) avoids this problem.  We see that even if all the agents have the same discount factor, then the price-dividend ratio will depend on the beliefs of all the agents.  Furthermore, the volatility of the stock price will be affected by the beliefs of the agents.    

\subsection{The Interest Rate Process}
We will now derive an expression for the riskless rate.  Define 
\begin{align*}
L_{t}  \equiv \delta_{t}^{R} \zeta_{t}  = \sum_{|\beta| = R} \binom{R}{\beta} \exp\{ \dfrac{\alpha \cdot \beta}{R}  X_{t}  - \dfrac{\gamma \cdot \beta}{R}  - \left[\dfrac{\rho \cdot \beta}{R} + \dfrac{\alpha^{2} \cdot \beta}{2 R}  \right] t \}
\end{align*}
Performing an It\^{o} expansion on $L$ we obtain:
\begin{align*}
dL_{t} = L_{t} ( \bar{\alpha}_{t} dX_{t} - \bar{\rho}_{t} dt )
\end{align*}
where 
\begin{align*}
\bar{\alpha}_{t} &= L_{t}^{-1} \sum_{|\beta| = R} \binom{R}{\beta} \dfrac{ \alpha \cdot \beta}{R} \exp\{ \dfrac{\alpha \cdot \beta}{R}  X_{t}  - \dfrac{\gamma \cdot \beta}{R}  - \left[\dfrac{\rho \cdot \beta}{R} + \dfrac{\alpha^{2} \cdot \beta}{2 R}  \right] t \} \\
\bar{\rho}_{t} &= L_{t}^{-1} \sum_{|\beta| = R} \binom{R}{\beta} \left( \dfrac{\rho \cdot \beta}{R} + \dfrac{\alpha^{2} \cdot \beta}{2 R} - \dfrac{1}{2} \left(\dfrac{\alpha \cdot \beta}{R}  \right)^{2} \right) \exp\{ \dfrac{\alpha \cdot \beta}{R}  X_{t}  - \dfrac{\gamma \cdot \beta}{R}  - \left[\dfrac{\rho \cdot \beta}{R} + \dfrac{\alpha^{2} \cdot \beta}{2 R}  \right] t \}
\end{align*}
An It\^{o} expansion on $\zeta_{t} = L_{t} \delta_{t}^{-R}$ then gives:
\begin{align*}
d \zeta_{t} = \zeta_{t} ( - r_{t} dt - \kappa_{t} dX_{t} ) 
\end{align*} 
where 
\begin{align*}
r_{t} = \bar{\rho}_{t} + R \sigma ( \alpha^{*} + \bar{\alpha}_{t}) - \dfrac{\sigma^{2} R(R+1)}{2}, \qquad \kappa_{t} = R \sigma - \bar{\alpha}_{t}
\end{align*}

\paragraph{Remarks}

We can now see how the riskless rate depends on the risk aversion of the agents.  We see that for $R> \sigma^{-1} (\alpha^{*} + \bar{\alpha}_{t} - \dfrac{\sigma}{2})$, the riskless rate is decreasing in $R$.  This is as we would expect, since more risk averse agents will need less encouragement to put their wealth into the risky asset.  

\subsection{The Stock Price Dynamics}
We will now explore the volatility and drift of the stock.  First define:
\begin{align*}
Z_{t} =\sum_{|\beta| = R} \frac{ \binom{R}{\beta} \exp\{ \dfrac{\alpha \cdot \beta}{R}  X_{t}  - \dfrac{\gamma \cdot \beta}{R}  - \left[\dfrac{\rho \cdot \beta}{R} + \dfrac{\alpha^{2} \cdot \beta}{2 R}  \right] t \} }{\dfrac{ \rho \cdot \beta}{R} + \dfrac{ \alpha^{2} \cdot \beta}{2 R} + \left(\dfrac{\sigma^{2}}{2} - \alpha^{*} \sigma\right) (1-R) - \dfrac{1}{2} \left(\dfrac{ \alpha \cdot \beta}{R} + (1-R) \sigma \right)^{2}  }
\end{align*}
We may then deduce that:
\begin{align*}
dZ_{t} = Z_{t} ( \tilde{\alpha}_{t} dX_{t} - \tilde{\rho}_{t} dt )
\end{align*}
where
\begin{align*}
\tilde{\alpha}_{t} & = Z_{t}^{-1} \sum_{|\beta| = R} \frac{ \binom{R}{\beta} \dfrac{\alpha \cdot \beta}{R} \exp\{ \dfrac{\alpha \cdot \beta}{R}  X_{t}  - \dfrac{\gamma \cdot \beta}{R}  - \left[\dfrac{\rho \cdot \beta}{R} + \dfrac{\alpha^{2} \cdot \beta}{2 R}  \right] t \} }{\dfrac{ \rho \cdot \beta}{R} + \dfrac{ \alpha^{2} \cdot \beta}{2 R} + \left(\dfrac{\sigma^{2}}{2} - \alpha^{*} \sigma\right) (1-R) - \dfrac{1}{2} \left(\dfrac{ \alpha \cdot \beta}{R} + (1-R) \sigma \right)^{2}  }  \\
\tilde{\rho}_{t} & = Z_{t}^{-1} \sum_{|\beta| = R} \frac{ \binom{R}{\beta}  \left( \dfrac{\rho \cdot \beta}{R} + \dfrac{\alpha^{2} \cdot \beta}{2 R} - \dfrac{1}{2} \left(\dfrac{\alpha \cdot \beta}{R}  \right)^{2} \right)           \exp\{ \dfrac{\alpha \cdot \beta}{R}  X_{t}  - \dfrac{\gamma \cdot \beta}{R}  - \left[\dfrac{\rho \cdot \beta}{R} + \dfrac{\alpha^{2} \cdot \beta}{2 R}  \right] t \} }{\dfrac{ \rho \cdot \beta}{R} + \dfrac{ \alpha^{2} \cdot \beta}{2 R} + \left(\dfrac{\sigma^{2}}{2} - \alpha^{*} \sigma\right) (1-R) - \dfrac{1}{2} \left(\dfrac{ \alpha \cdot \beta}{R} + (1-R) \sigma \right)^{2}  } 
\end{align*}
Using expression (\ref{CRRASP1}), we may then deduce that the stock price process satisfies:
\begin{align*}
dS_{t} = S_{t} \{ (\sigma + \tilde{\alpha}_{t} - \bar{\alpha}_{t}) dX_{t} + ( \bar{\rho}_{t} - \tilde{\rho}_{t} + \sigma \alpha^{*} + (\tilde{\alpha}-\bar{\alpha})(\sigma - \bar{\alpha}) ) dt  \}
\end{align*}
In particular, we may note that the volatility is given by:
\begin{align*}
\sigma_{t}^{S} = \sigma + \tilde{\alpha}_{t} - \bar{\alpha}_{t}
\end{align*}

\paragraph{Remark}
As mentioned earlier, even if the agents all have the same discount factors, the volatility of the stock price process will be different to the volatility of the dividend process. This is in contrast to the case of logarithmic agents.  The model presented here is therefore more likely to agree with data, since in general the volatility of the dividend process and stock price process will be different. 

\subsection{Portfolios and Volume of Trade}

In order to understand the volume of trade, we must first understand the portfolio held by the different agents.  Note that the wealth of agent $j$ must satisfy:
\begin{align}
dw_{t}^{j} = \pi_{t}^{j} (dS_{t} + \delta_{t}) - c_{t}^{j} dt + (w_{t}^{j} - \pi_{t}^{j} S_{t}) r_{t}  \label{CRRAWealthBudgetEquation}
\end{align}
where $\pi_{t}^{j}$ denotes the amount of the risky asset held by agent agent $j$ at time $t$.  But we also know that the wealth of agent $j$ is given by equation (\ref{CRRAWealth}).  If we perform an It\^{o} expansion on equation (\ref{CRRAWealth}), then we may compare with equation (\ref{CRRAWealthBudgetEquation}) and read off the portfolio held by agent $j$.

Proceeding in this way, first let
\begin{align*}
Z_{t}^{j} =  \sum_{|\beta| = R-1} \frac{\binom{R-1}{\beta} \exp\{ \dfrac{\alpha_{j} + \alpha \cdot \beta}{R}  X_{t}  - \dfrac{\gamma_{j} + \gamma \cdot \beta}{R}  - \left[\dfrac{\rho_{j} + \rho \cdot \beta}{R} + \dfrac{\alpha_{j}^{2} + \alpha^{2} \cdot \beta}{2 R}  \right] t \} }{  \left[\dfrac{\rho_{j} + \rho \cdot \beta}{R} + \dfrac{\alpha_{j}^{2} + \alpha^{2} \cdot \beta}{2 R} + \left(\dfrac{\sigma^{2}}{2} - \alpha^{*} \sigma\right) (1-R) - \dfrac{1}{2} \left(\dfrac{\alpha_{j} + \alpha \cdot \beta}{R} + (1-R) \sigma \right)^{2}  \right] }
\end{align*}
Then we may write:
\begin{align*}
dZ_{t}^{j} = Z_{t}^{j} (\tilde{\alpha}_{t}^{j} dX_{t} - \tilde{\rho}_{t}^{j} dt)
\end{align*}
where 
\begin{align*}
\tilde{\alpha}_{t}^{j} Z_{t}^{j} = \sum_{|\beta| = R-1} \frac{\binom{R-1}{\beta} \dfrac{\alpha_{j} + \alpha \cdot \beta}{R} \exp\{ \dfrac{\alpha_{j} + \alpha \cdot \beta}{R}  X_{t}  - \dfrac{\gamma_{j} + \gamma \cdot \beta}{R}  - \left[\dfrac{\rho_{j} + \rho \cdot \beta}{R} + \dfrac{\alpha_{j}^{2} + \alpha^{2} \cdot \beta}{2 R}  \right] t \} }{  \left[\dfrac{\rho_{j} + \rho \cdot \beta}{R} + \dfrac{\alpha_{j}^{2} + \alpha^{2} \cdot \beta}{2 R} + \left(\dfrac{\sigma^{2}}{2} - \alpha^{*} \sigma\right) (1-R) - \dfrac{1}{2} \left(\dfrac{\alpha_{j} + \alpha \cdot \beta}{R} + (1-R) \sigma \right)^{2}  \right] }
\end{align*}
and $\tilde{\rho}_{t}^{j}$ is another process, not currently of interest to us.  By It\^{o}'s formula:
\begin{align}
dw_{t}^{j} = w_{t}^{j} \{ (\sigma + \tilde{\alpha}_{t}^{j} - \bar{\alpha}_{t}) dX_{t} + ( \bar{\rho}_{t} - \tilde{\rho}_{t}^{j} + \sigma \alpha^{*} + ( \tilde{\alpha}_{t}^{j} - \bar{\alpha}_{t})( \sigma - \bar{\alpha}_{t})  \}  \label{CRRAWealthItoExpansion}
\end{align}
Comparing (\ref{CRRAWealthItoExpansion}) with (\ref{CRRAWealthBudgetEquation}), we obtain the proportion of risky asset held by agent $j$ as:
\begin{align}
\pi_{t}^{j} = \frac{ w_{t}^{j} ( \sigma + \tilde{\alpha}_{t}^{j} - \bar{\alpha}_{t})}{ S_{t} ( \sigma + \tilde{\alpha}_{t} - \bar{\alpha}_{t})}  \label{PropRiskyAsset}
\end{align}
In exactly the same manner as \citet{BrownRogersDiverseBeliefs}, we may then perform an It\^{o} expansion on $\pi_{t}^{j}$.  We may then interpret the quadratic variation of $\pi_{t}^{j}$ as the volume of trade of agent $j$.  The calculation is omitted here, because it is straightforward and the analysis adds little to our understanding.  Future work could explore the behaviour of this volume of trade.

\section{Conclusion}

We have explored a model in which agents all have diverse beliefs and have CRRA utility.  We showed how the analysis is just as simple as in the case of logarithmic utility; the only complication is the inclusion of a multinomial sum.  In particular, we derived expressions for the state price density, wealth processes, the stock price and the riskless rate.  We also explored the stock price dynamics and the portfolios of the agents.   We also showed how some of the undesirable properties of the logarithmic case are no longer present in the case of general CRRA agents.  In particular, we showed that the agents could have the same discount factor and the diverse beliefs would have an impact on the stock price volatility.   

The purpose of this note was simply to illustrate the ease with which we can solve a diverse beliefs equilibrium with CRRA agents and integer coefficient of relative risk aversion.  Future work should include a numerical study of the behaviour of the quantities studied in this note.  In particular, there is the possibility of fitting the model to stock market data, as in an appendix of \citet{BrownRogersDiverseBeliefs}.  The extra degree of freedom given by the risk aversion parameter could allow a better match to the data.

\bibliography{references}
\bibliographystyle{jmb}

\end{document}